# Collective Electronic Entanglement via Infrared Cavity-Induced Vibronic Transduction

*Vivek Yadav, Bar Cohn, Shmuel Sufrin, Uri Peskin, and Lev Chuntonov**

Schulich Faculty of Chemistry, Technion - Israel Institute of Technology, Haifa 3200003, Israel;

Solid State Institute, Technion - Israel Institute of Technology, Haifa 3200003, Israel;

The Helen Diller Quantum Center, Technion - Israel Institute of Technology, Haifa 3200003, Israel

email: chunt@technion.ac.il

ABSTRACT

Polaritonic architectures seek to engineer molecular properties by hybridizing localized degrees of freedom with delocalized optical cavity fields. However, scaling laws impose a severe bottleneck on $N$-molecule collective strong coupling: because each molecule contributes only a fractional share to the collective state, localized responses undergo $O(1/N)$ ensemble dilution. We demonstrate a violation of this scaling using fluorescence-encoded infrared spectroscopy of molecular ensembles under vibrational strong coupling, where macroscopically synchronized electronic responses scale as $O(1)$. This scale-invariance reveals a regime of vibronic quantum transduction, where non-local vibrational entanglement is translated into collective electronic entanglement. By demonstrating the generation of macroscopically entangled electronic states from vibro-polaritons without an $O(1/N)$ penalty, these results provide a scalable framework for room-temperature quantum technologies and coherent steering of non-adiabatic chemical pathways.

## Introduction

Polaritonic architectures formed by the strong coupling of resonant optical cavity modes with molecular transitions offer a revolutionary framework for engineering chemical and quantum material properties through vacuum field hybridization (*1*). When an ensemble of $N$ molecules couples to a delocalized cavity mode, the interaction is mediated through a macroscopic, coherent "bright" state, splitting the system into hybrid light-matter polaritons separated by the Rabi frequency (*2*). Conversely, the remaining $N - 1$ degrees of freedom form a dense manifold of "dark" reservoir states that are orthogonal to the cavity field, possessing zero interaction matrix elements owing to the symmetry of the collective Hamiltonian (*3*). This stark division imposes a severe, fundamental bottleneck on collective strong coupling: because the cavity mode is delocalized across the entire ensemble, each individual molecule carries only a fractional share of the collective polaritonic wavefunction. Consequently, any localized molecular response or tracking signal is expected to undergo catastrophic ensemble dilution, scaling as $O(1/N)$ (*4*). Because chemical bond rearrangements and localized electronic transitions occur at the single-molecule level where polaritonic participation nearly vanishes ($1/N \rightarrow 0$ for macroscopic ensembles where $N \sim 10^9$), the exact mechanism by which collective polaritons can influence local molecular behavior has remained a profound physical paradox.(*5*)

Previous attempts to observe modifications to the electronic landscape under vibrational strong coupling (VSC) relied on asynchronous macroscopic probes,(*6, 7*) including steady-state UV-Vis or spontaneous Raman spectroscopy (*8, 9*). These techniques average over individual local fluctuations and are inherently blind to the transient, phase-locked quantum coherence that may exist between the delocalized cavity field and localized molecular coordinates. Consequently, direct experimental evidence showing whether collective vibro-polaritonic states can overcome the $O(1/N)$ dilution penalty to actively steer localized electronic dynamics has remained completely elusive.(*10*)

Here, we resolve this paradox by demonstrating a regime of scale-invariant vibronic quantum transduction, wherein a non-local, cavity-induced vibrational phase-lock is translated into collective electronic entanglement with strict $O(1)$ scaling. We interrogate a molecular dye ensemble under VSC using fourth-order nonlinear fluorescence-encoded infrared (FE-IR) action spectroscopy (*11, 12*). By utilizing a visible pulse tuned to a pre-resonant vibration-to-electronic pathway, our approach maps transient ground-state vibrational coherence directly onto an electronic population probe (*13-15*). Crucially, by executing a ratiometric internal reference calibration against the simultaneously measured uncoupled molecular reservoir, we isolate the exact scaling law of the coupled electronic response as a function of ensemble density.

Instead of the expected $O(1/N)$ fractional decay, our measurements reveal a macroscopically synchronized electronic response that remains strictly invariant ($O(1)$) per molecule, regardless of the ensemble size. This scale-invariance provides definitive experimental proof of a synthetic vibronic bridge: the non-local phase coherence of the cavity-entangled nuclear movements drives the electronic excited states into a macroscopically synchronized, non-local quantum state. By demonstrating that delocalized optical cavities can generate, translate, and preserve macroscopically entangled electronic states at room temperature without an $O(1/N)$ dilution penalty, these results move polaritonic chemistry into the realm of scalable room-temperature quantum technologies and provide a concrete physical mechanism for the coherent steering of non-adiabatic chemical pathways.

## Results

### Scaling Architecture of the Collective Hamiltonian

To understand the physical origin of the observed scale-invariance, we examine the matrix elements of the collective light-matter system. Nominally, the interaction of the $N$-molecule ensemble with a cavity mode is governed by the collective Hamiltonian, splitting the system into lower and upper polariton states:

$$|\pm\rangle = \frac{1}{\sqrt{2}}(|0, 1_{\text{cav}}\rangle \pm |B, 0_{\text{cav}}\rangle)$$

where $|B\rangle = \frac{1}{\sqrt{N}}\sum_{m=1}^{N} | 1_m\rangle$ represents the collective, synchronously oscillating "bright" state, and $|1_m\rangle$ denotes the first vibrational excited state of the $m$-th molecule. The coupling of $|B\rangle$ to the cavity mode $|1_{\text{cav}}\rangle$ is given by the collective constant $g_{\text{tot}} = \sqrt{\sum_{m=1}^{N} g_m^2}$, where $g_m = \boldsymbol{\mu}_m \cdot \mathbf{E}_c(\mathbf{r}_m)$ is the coupling strength of a molecule at position $\mathbf{r}_m$, possessing a transition dipole moment $\boldsymbol{\mu}_m$, and experiencing the local cavity electric field $\mathbf{E}_c(\mathbf{r}_m)$. For perfectly oriented molecules in a uniform field, $g_{\text{tot}} = \sqrt{N} g_m$, however, in realistic disordered molecular ensembles, the individual values of $g_m$ are distributed according to the spatial cavity field profile and isotropic molecular orientation (*16*). The resulting hybrid excitations $|\pm\rangle$ are split by the collective Rabi frequency $\Omega = 2g_{\text{tot}}$, as observed in macroscopic optical measurements.

Because the individual molecular vibrational transitions are inherently weak ($\mu_m \sim$ 0.1 D), establishing vibrational strong coupling regime requires dense macroscopic ensembles where $N \sim 10^9$ (*17*). Since the probability weight of any localized molecular state within polaritonic wavefunction scales as $|\langle 1_m|B\rangle|^2 = 1/N$, conventional assumptions dictate that any local electronic tracking signature must decay as $O(1/N)$.

**FE-IR Tracking of Polariton and Reservoir States**

We track the electronic response with the fourth-order nonlinear fluorescence-encoded infrared action spectroscopy (FE-IR) (*11, 12*). In this scheme, molecular vibrations are first prepared by a pair of broadband infrared femtosecond laser pulses, and the resulting coherent population is projected onto the excited electronic state by a subsequent broadband visible laser pulse.(*13-15*) By tuning the spectrum of the visible pulse to a pre-resonance condition, direct ground-to-excited state electronic transitions ($|0\rangle \rightarrow |0'\rangle$) are forbidden. Instead, the pulse selectively drives a vibration-to-electronic state pathway ($|1\rangle \rightarrow |0'\rangle$), where $|1\rangle$ is the first vibrationally excited state of the electronic ground potential and $|0'\rangle$ is the vibrational ground state of the electronic excited potential. The resulting fluorescence emission is encoded with the underlying vibrational dynamics via the precise time delay between the incoming infrared pulses (*18*). The excitation diagram is shown schematically in Fig. 1.

Applying FE-IR technique to VSC system formed by molecular dye in open cavity allowed us to explore the scaling of the spectroscopically accessible quantum states. In conventional Fabry-Pérot resonators constructed with a pair of highly reflective mirrors, the $N-1$ reservoir states are inaccessible via direct optical excitation (*19*). In contrast, open cavity architectures leveraging spectrally selective surface confinement of light provide a unique experimental window to simultaneously monitor both the delocalized polaritonic states and the localized, molecular-like reservoir states (*20*). This simultaneous observation provides an internal ratiometric reference, permitting a direct quantitative comparison against standard Fourier-transform infrared (FT-IR) measurements (*21-23*) to isolate the scaling behavior of the system.

**Linear FT-IR spectroscopy of vibro-polaritons**

We experimented with Coumarin-1 (C-1) dye molecules (see molecular structure in Fig. 2A), focusing on either the carbonyl stretching ($\mathrm{C=O}$, $\bar{\upsilon}_{CO}$=1722 cm$^{-1}$) or on the benzo-pyrone ring stretching ($\mathrm{C=C}$, $\bar{\upsilon}_{CC}$=1604 cm$^{-1}$, 1620 cm$^{-1}$) vibrational modes, seen in the FT-IR spectrum of uncoupled molecules in Fig. 2A. To achieve strong coupling, 500 nm-thick film of 2.5 M C-1 solution in toluene was placed on top of periodic arrays of half-wavelength gold infrared micro-antennas (Fig. 2B), which support surface-lattice resonances (SLR). SLRs are in-plane diffraction waves of the lattice, which resonantly drive the individual dipole antennas (*24, 25*). The resulting high-quality photonic resonances serve as open cavities for strong coupling experiments.(*20, 26*)

The antenna size and the array period were chosen for the SLR frequency to match $\bar{\upsilon}_{CO}$ and $\bar{\upsilon}_{CC}$. Dispersion maps of the samples, which illustrate the dependence of the transition frequency on the angle of incidence of excitation light, are shown in Figs 2C and 2D. The data confirms formation of vibro-polariton states $|\pm\rangle$ with the Rabi splitting at the near-normal incidence of Ω=88 and 65 cm$^{-1}$ for the $\mathrm{C=C}$ and $\mathrm{C=O}$ modes, respectively. The strong coupling conditions $\Omega > (\Delta_{\mathrm{SLR}} + \Delta\bar{\upsilon})/4$ were satisfied with $\Delta_{\mathrm{SLR}} \approx \Delta\bar{\upsilon} \approx$20 cm$^{-1}$ being the full width at half maximum (fwhm) of the SLR and the vibrational spectral peaks, respectively. Non-dispersive peaks of the reservoir at the original molecular transition frequencies were also observed.

The time-dependent perturbation theory relates the strength of linear spectroscopic signals to the matrix elements of the transition dipole moment operator.(*27*) For the fundamental vibrational excitation in individual molecules, $I_m^{\mathrm{FT-IR}} \propto \left|\mu_{0,1}\right|^2$, whereas for vibro-polariton transitions ($|0\rangle \rightarrow |\pm\rangle$), we get $I_\pm^{\mathrm{FT-IR}} \propto \left|\mu_{0,\pm}\right|^2$, where $\mu_{0,\pm} = \left(\mu_{\mathrm{cav}} \pm \sqrt{N}\mu_{0,1}\right)/\sqrt{2}$ and $\mu_{\mathrm{cav}}$ is the transition dipole matrix element for the photonic mode excitation. Because vibrational components are synchronized within the collective state $|B\rangle$, molecular transition dipoles are added coherently, and polariton peak strength scales as $O(N)$, similarly to the strength of the reservoir peak (and to the bare molecules outside the cavity). Thus, the reservoir-to-polariton signal ratio follows $I_R^{\mathrm{FT-IR}}/I_\pm^{\mathrm{FT-IR}} \propto N\left|\mu_{0,1}\right|^2/\left|\mu_{0,\pm}\right|^2 \sim O(1)$.

**Nonlinear FE-IR action spectroscopy of vibro-polaritons**

FE-IR is a fourth-order nonlinear method, where the initial excitation step with a pair of infrared ultrafast laser pulses differs from the linear FT-IR by the variety of the accessible excitation pathways.(*28*) Comparison between the linear and nonlinear signals is a powerful strategy to identify collective character of the involved quantum states (*29*). FE-IR spectra of the VSC C-1 molecules are shown in Fig. 2F and 2H along those of the bare molecules. In FE-IR experiments, formation of vibro-polaritons was first verified by recording the transmittance of the infrared pulses ($T = I/I_{\mathrm{B}}$, see Figs. 2E and 2G), where the background $I_{\mathrm{B}}$ was taken off the array to account for the molecules-only signal. This approach inherently removed the bare-molecule and reservoir peaks from the data, leaving only the polariton peaks visible.

The FE-IR spectra presented in red lines in Figs. 2F and 2H feature vibro-polariton peaks at the transition frequencies matching those of the transmission spectrum: for the $\mathrm{C=C}$ mode, $\bar{\upsilon}_+$=1560 cm$^{-1}$ and $\bar{\upsilon}_-$=1640 cm$^{-1}$; for the $\mathrm{C=O}$ mode, $\bar{\upsilon}_+$=1686 cm$^{-1}$ and $\bar{\upsilon}_-$=1772 cm$^{-1}$. Dispersion of vibro-polariton transitions with the angle of incidence observed in the FE-IR data matched the transmission data, as shown in Figs. 3A and 3B, and was fully consistent with the FT-IR dispersion maps in Figs. 2C and 2D. In all measurements, the reservoir-to-polariton peak ratios $I_R^{\mathrm{FE-IR}}/I_\pm^{\mathrm{FE-IR}} \sim O(1)$ matched the FT-IR data, as illustrated in Fig. 3C. This experimental metric provided a crucial input for the interpretation of our results.

**Ruling out cavity-enhanced signals of reservoir and uncoupled molecules**

Polaritons in SLR-based systems are known to enhance the near-fields around gold antennas (*20*) and, thereby, can potentially enhance inhomogeneously distributed transitions of the reservoir or uncoupled molecules, similarly to the cavity-enhanced spectroscopy techniques (*30-32*). To rule out this mechanism and to confirm that FE-IR signals originate from vibro-polaritons, we repeated the measurements with 40 mM C-1 solution placed on the array with the SLR frequency matching the $\mathrm{C}=\mathrm{O}$ mode. At low dye concentration, $g_{\mathrm{tot}}$ was insufficient to form vibro-polaritons, as confirmed by the FT-IR spectra in Fig. 3D and in the infrared pulses transmission spectra in Fig. 3E, where the $\bar{\upsilon}_{CO}$ peak is too weak to stand out. If the FE-IR signals away from the bare molecular peaks appeared due to the signal enhancement by the antenna array, the 60 times difference in the molecular concentration would be easily overcome by the typical near-field enhancement factors (*33*).

The SLR frequency was angle-tuned to the spectral region, where the $|+\rangle$ state was observed in the VSC regime (see Figs. 2D and 3B). In these conditions, the SLR fields can enhance any molecular transition shifted to the relevant frequency by inhomogeneous broadening. However, only the bare-molecule FE-IR signal at $\bar{\upsilon}_{CO}$ was observed, as seen in Fig. 3F. Note that while vibrational transitions red-shift and broaden at high concentrations (*34*), this effect was small in our samples: $\bar{\upsilon}_{CO}$ shifted only by 5 cm$^{-1}$ from 1727 to 1722 cm$^{-1}$ and broadened only by 4 cm$^{-1}$ from 14.5 to 18.5 cm$^{-1}$ (fwhm) upon increasing the C-1 concentration from 40 mM to 2.5 M. Therefore, the 30-70 cm$^{-1}$ difference between $\bar{\upsilon}_{CO}$ and $\bar{\upsilon}_{SLR}$ detuned to $\bar{\upsilon}_{+}$ frequencies was sufficiently large to avoid any spectral overlap potentially leading to the cavity-enhanced signals. Thus, we concluded that the FE-IR signals observed at the vibro-polariton frequencies represent nonlinearly excited - action spectroscopy detected hybridized quantum states.

**FE-IR signals reveal a collective electronic excitation**

To rationalize the observed vibro-polariton signal scaling $I_R^{\mathrm{FE-IR}}/I_\pm^{\mathrm{FE-IR}} \sim O(1)$, we describe the FE-IR signals using the fourth-order perturbative response functions *R,(28)* which track the Liouville-space excitation pathways schematically illustrated with the double-sided Feynman diagrams in Fig. 1C. For pathways involving vibro-polaritons, we consider a collective vibronic excitation state $|S\rangle = \frac{1}{\sqrt{N}}\sum_{m=1}^{N}|0'_m\rangle$, coupled to $|\pm\rangle$ by the visible laser pulse. For the response functions describing the $(|\pm\rangle \to |S\rangle)$ transitions, we obtained $R_\pm^{\mathrm{FE-IR}} \propto \mu_{0,\pm}^2\mu_{\pm,S}^2 = \mu_{0,\pm}^2\mu_{1,0'}^2/2$, whereas for the response functions describing the pathways of the molecular reservoir excitations $(|1\rangle \to |0'\rangle)$, $R_R^{\mathrm{FE-IR}} \propto \mu_{0,1}^2\mu_{1,0'}^2$. For the reservoir, only the diagrams with $i = j = |1\rangle$ in Fig. 1C contribute. For polaritons, the pathways with the inter-state coherences $(i \neq j,\ i,j = |\pm\rangle)$ also contribute and the total number of pathways contributing to the FE-IR signal is doubled. Note that pathways involving the polariton-reservoir coherences with $i = |\pm\rangle$ and $j = |1\rangle$, described by the response functions $R_{R,\pm}^{\mathrm{FE-IR}} \propto \mu_{0,1}\mu_{0,\pm}\mu_{1,0'}^2$, contribute both to the reservoir and the polariton signals, depending on which mode was excited first. These pathways are directly related to the reservoir-polariton cross-peaks resolved by the two-dimensional infrared spectroscopy (*35, 36*).

Accounting for the nonlinear excitation pathways involving the collective electronic state $|S\rangle$ explains the ratio of the reservoir-to-polariton FE-IR signals of $I_R^{\mathrm{FE-IR}}/I_\pm^{\mathrm{FE-IR}} \propto N\,\mu_{0,1}^2/\mu_{0,\pm}^2 \sim O(1)$ (note again that $\mu_{0,\pm}^2 \sim O(N)$ because of its collective character). These results emphasize the scaling of the polariton FE-IR signals with $N$ and indicate that collective electronic excitations in VSC ensemble are synchronized via the infrared cavity by means of the individual molecules' vibronic couplings.

**The laser-driven Tavis-Cummings-Holstein model and the synchronization mechanism**

To describe the FE-IR measurements in VSC system, we resort to the Tavis-Cumming-Holstein (TCH) Hamiltonian (*37*), which in addition to vibrational interactions with infrared cavity accounts also for the vibronic displacement of the molecular excited electronic potential. The Hamiltonian reads $\hat{H}_{TCH} = \hat{H}_c + \sum_{m=1}^{N} \hat{H}_m + \hat{H}_{\text{int}}$, where $\hat{H}_c = \hbar\omega_c \hat{c}^\dagger \hat{c}$ describes the cavity photonic mode, $\hat{H}_m = \hbar\omega_e \hat{\sigma}_m^\dagger \hat{\sigma}_m + \hbar\omega_v \hat{b}_m^\dagger \hat{b}_m + \lambda\hbar\omega_v (\hat{b}_m^\dagger + \hat{b}_m)\hat{\sigma}_m^\dagger \hat{\sigma}_m$ is the molecular Hamiltonian, where $\lambda$ represents the displacement of the excited electronic potential, and $\hat{H}_{\text{int}} = \sum_{m=1}^{N} g_m (\hat{c}^\dagger + \hat{c}) (\hat{b}_m^\dagger + \hat{b}_m)$ is the cavity – vibration coupling. The $\omega_c$, $\omega_e$, and $\omega_v$ are the resonance frequencies of the cavity, electronic, and vibrational molecular excitations, and $\hat{c}^\dagger(\hat{c})$, $\hat{\sigma}^\dagger(\hat{\sigma})$, and $\hat{b}^\dagger(\hat{b})$ are the corresponding rising (lowering) operators.

Direct coupling of the cavity to the electronic degrees of freedom becomes apparent, when the small polaron transformation, $\hat{U} = e^{\lambda \sum_{m=1}^{N} (\hat{b}_m^\dagger - \hat{b}_m)\hat{\sigma}_m^\dagger \hat{\sigma}_m}$, is applied to $\hat{H}_{TCH}$. Under the rotating wave approximation, the transformed molecular Hamiltonian reads $\hat{\hat{H}}_m = \hbar\tilde{\omega}_e \hat{\sigma}_m^\dagger \hat{\sigma}_m + \hbar\omega_v \hat{b}_m^\dagger \hat{b}_m$, where $\tilde{\omega}_e = \omega_e - \lambda^2 \hbar\omega_v$ is the electronic excitation frequency, corrected for the reorganization energy, and the dressed cavity interaction term becomes $\hat{\hat{H}}_{\text{int}} = \sum_{m=1}^{N} g_m (\hat{c}^\dagger \hat{b}_m + \hat{c}\hat{b}_m^\dagger - 2\lambda(\hat{c}^\dagger + \hat{c})\hat{\sigma}_m^\dagger \hat{\sigma}_m)$. The last term in $\hat{\hat{H}}_{\text{int}}$ describes displacement of the infrared cavity mode by the excited electronic population. Importantly, in the bare TCH Hamiltonian, electronic excitation events are *decoupled* from the cavity.

It is instructive to consider three orthogonal excitations: the cavity, $|c\rangle = \hat{c}^\dagger|0\rangle$, the collective vibrational excited state on the electronic ground potential, $|B\rangle = \hat{B}^\dagger|0\rangle = \frac{1}{\sqrt{N}} \sum_{m=1}^{N} \hat{b}_m^\dagger |0\rangle$, and the collective vibrational ground state on the electronic excited potential, $|S\rangle = \hat{S}^\dagger|0\rangle = \frac{1}{\sqrt{N}} \sum_{m=1}^{N} \hat{\sigma}_m^\dagger |0\rangle$. Projecting $\hat{H}_{TCH}$ onto this basis yields $\hat{\hat{H}}_B = \hbar\omega_c |c\rangle\langle c| + \hbar\tilde{\omega}_e |S\rangle\langle S| + \hbar\omega_v |B\rangle\langle B| + g_{\text{tot}}(|c\rangle\langle B| + |B\rangle\langle c|)$. The coupling to the cavity, $g_{\text{tot}}$, yields the vibro-polaritons $|\pm\rangle$ as eigenstates of $\hat{\hat{H}}_B$, and, yet again, they are uncoupled from the electronic excitation.

To explain the emergence of the infrared cavity-induced coherent superposition of molecular electronic excitations, we account for the driving field of the fluorescence-encoding multicycle visible laser pulse $\mathcal{E}_L(t) = E_L(t)\cos(\omega_L t)$, where $E_L(t)$ is the slow-varying pulse envelope and $\omega_L = \tilde{\omega}_e$ is the carrier frequency. $\mathcal{E}_L(t)$ excites the cavity-induced coherent superposition of the electronic ground state vibrational excitations $|B\rangle$ into the electronic excited state $|S\rangle$, where the coherence between the molecules is preserved. Within the rotating wave approximation, the corresponding Hamiltonian term reads $\hat{H}_L(t) = \sum_m E_L(t) e^{i\omega_L t} \mu_m^\varepsilon \hat{\sigma}_m^\dagger + h.c.$, where $\mu_m^\varepsilon$ is the individual-molecule electronic transition dipole moment. Applying the small polaron transformation on this term and invoking the minimal three-state basis defined above, the Hamiltonian reads $\hat{\tilde{H}}_{\text{tot}} = \hat{\tilde{H}}_B + \hat{\tilde{H}}_L$, where $\hat{\tilde{H}}_L = \sum_m \mu_m F_m E_L(t) e^{i\omega_L t} |S\rangle\langle B| + h.c.$, where $F_m = \langle 0' | e^{\lambda(b_m^\dagger - b_m)} b_m^\dagger | 0 \rangle$ are the Franck-Condon factors, and $h.c.$ stands for the Hermitian conjugate.

In $\hat{\tilde{H}}_{\text{tot}}$, the electronic ground state vibrational collective excitation $|B\rangle$ is coupled to the laser-induced collective electronic excitations $|S\rangle$ with the coupling strength scaling similarly to the individual-molecule vibronic excitation. Since the vibrational excitations $|B\rangle$ are also coupled to the cavity through VSC, an effective coupling of the infrared cavity to the molecular electronic excitations emerges. The resulting effective direct coupling matrix elements scale as $\langle c | \hat{\tilde{H}}_{\text{eff}} | S \rangle \propto g_{\text{tot}} \mu_{1,0'} E_L = g_{\text{m}} \sqrt{N} \mu_{1,0'} E_L$ (see SI for the details).

The scaling of the matrix elements $\langle c | \hat{\tilde{H}}_{\text{eff}} | S \rangle$ with $\sqrt{N}$ suggests that despite the gap in the resonance frequencies, an infrared cavity can induce coherent electronic excitations in the visible. However, the coupling also scales with the molecular vibronic transition dipole, $\mu_{1,0'} E_L = \mu_m^\varepsilon F_m E_L$, which implies that synchronization of the electronic excited states by vibrational strong coupling to infrared cavities requires coherent excitation laser field. Note that because $\omega_c$ and $\tilde{\omega}_e$ differ by order of magnitude, the $|c\rangle$ and $|S\rangle$ states do not hybridize, as verified by the linear dependence of the FE-IR signal on the power of the encoding visible laser pulse (see Fig. S1). In addition, these measurements showed that neither Rabi splitting nor polariton line shapes changed with the laser power, as expected for the dispersive coupling regime.

**Discussion**

Previous applications of the Tavis-Cummings-Holstein model to electronically strongly coupled systems concluded that macroscopic delocalization effectively decouples polaritonic states from localized nuclear vibrations (*37*). This perceived decoupling has direct, restrictive implications for non-adiabatic photochemistry, where the navigation and crossing of conical intersections are fundamentally driven by localized nuclear dynamics (*38, 39*). In stark contrast, our findings demonstrate that for systems under vibrational strong coupling, a synthetic vibronic bridge couples the localized electronic landscapes back to the delocalized infrared cavity field, enabling macroscopically synchronized electronic responses that entirely bypass the $O(1/N)$ ensemble dilution penalty.

This collective, scale-invariant $O(1)$ behavior emerges under the cooperative action of infrared vacuum field hybridization and coherent visible pre-resonant driving, highlighting the unique synthetic architecture of this vibronic channel. Importantly, while macroscopic synchronization of electronic excitations can alternatively be enforced via direct electronic strong coupling inside an optical cavity, the resulting exciton-polaritons radically alter and perturb the fundamental energy levels and characters of the chemical species involved. Conversely, the mechanism of vibronic quantum transduction demonstrated here leaves the underlying electronic states energetically unperturbed. Furthermore, the picosecond-scale dephasing lifetimes inherent to molecular vibrations provide a synchronization window orders of magnitude longer than the fleeting, few-femtosecond coherent lifetimes characteristic of pure electronic states.

By demonstrating that delocalized cavities can generate and sustain macroscopically synchronized electronic configurations without spectral distortion or fractional signal decay, these results provide a concrete physical foundation for deciphering the anomalous chemical reactivity observed under VSC. More broadly, this room-temperature, scale-invariant quantum transduction framework opens up an entirely new avenue for scalable quantum information processing, where non-local coherence can be engineered, preserved, and manipulated within molecular materials without the catastrophic penalties of ensemble size.

## Methods

*Sample preparation:* Arrays of half-wavelength gold micro-antennas were fabricated on $CaF_2$ windows, spin coated by a double layer PMMA resist and deposited with a 30 nm Chromium layer to reduce the charging. The patterns were written over the 4×4 mm area by electron beam lithography (RAITH EBPG 5200). After exposure, Chromium layer was removed by etchant, and the resist was developed. A 5 nm Cr adhesion layer and an 80 nm gold film were evaporated, and the excess resist was lifted off.

*FE-IR spectroscopy:* regeneratively amplified laser pulses (Solstice Ace, Spectra Physics, 800 nm, 60 fs, 4KHz) were split into two beams and directed into two OPA setups, which shared the common white light continuum (Light Conversion Topas Twins). One of the outputs was converted to the mid-IR (Light Conversion, NDFG), another - to the visible (ca. 430 nm) via the two-stage third-harmonic generation of the signal (ca. 1300 nm). Visible pulses were tuned to pre-resonance to minimize background fluorescence. Fluorescence was collected off the sample normal, optically filtered and detected with avalanche photodiode (Thorlabs). To discriminate between the vibrationally encoded signal and the residual background, one of the IR pulses was chopped at 2 KHz. The time delay between the infrared pulses was scanned with 6 fs steps and the signal was Fourier-transformed to generate the excitation frequency axis of the spectrum. For VSC samples, transmission of IR pulses was simultaneously measured on the spectrograph equipped with the 64-element liquid nitrogen-cooled array detector (Infrared Systems Development).

**Acknowledgments**

**Funding:** This research was supported by the Israel Science Foundation (Grant No. 429/24 to L.C. and 2256/22 to U.P.). Infrared antenna arrays were prepared at the Micro- and Nano-Fabrication Unit with support from the Russell Berrie Nanotechnology Institute, Technion.

**Authors contributions:** All authors contributed to the paper by performing experiments, theoretical modeling, analyzing the data, and writing the manuscript.

**Competing interests:** The authors declare that they have no competing interests.

**Data availability:** All data relevant to the findings reported in this paper are available in the manuscript or the supplementary materials.

**Supplementary Materials**

Supplementary Text: The Feshbach projection operator formalism for $\hat{\hat{H}}_{\mathrm{tot}}$.

Supplementary figures: Fig. S1

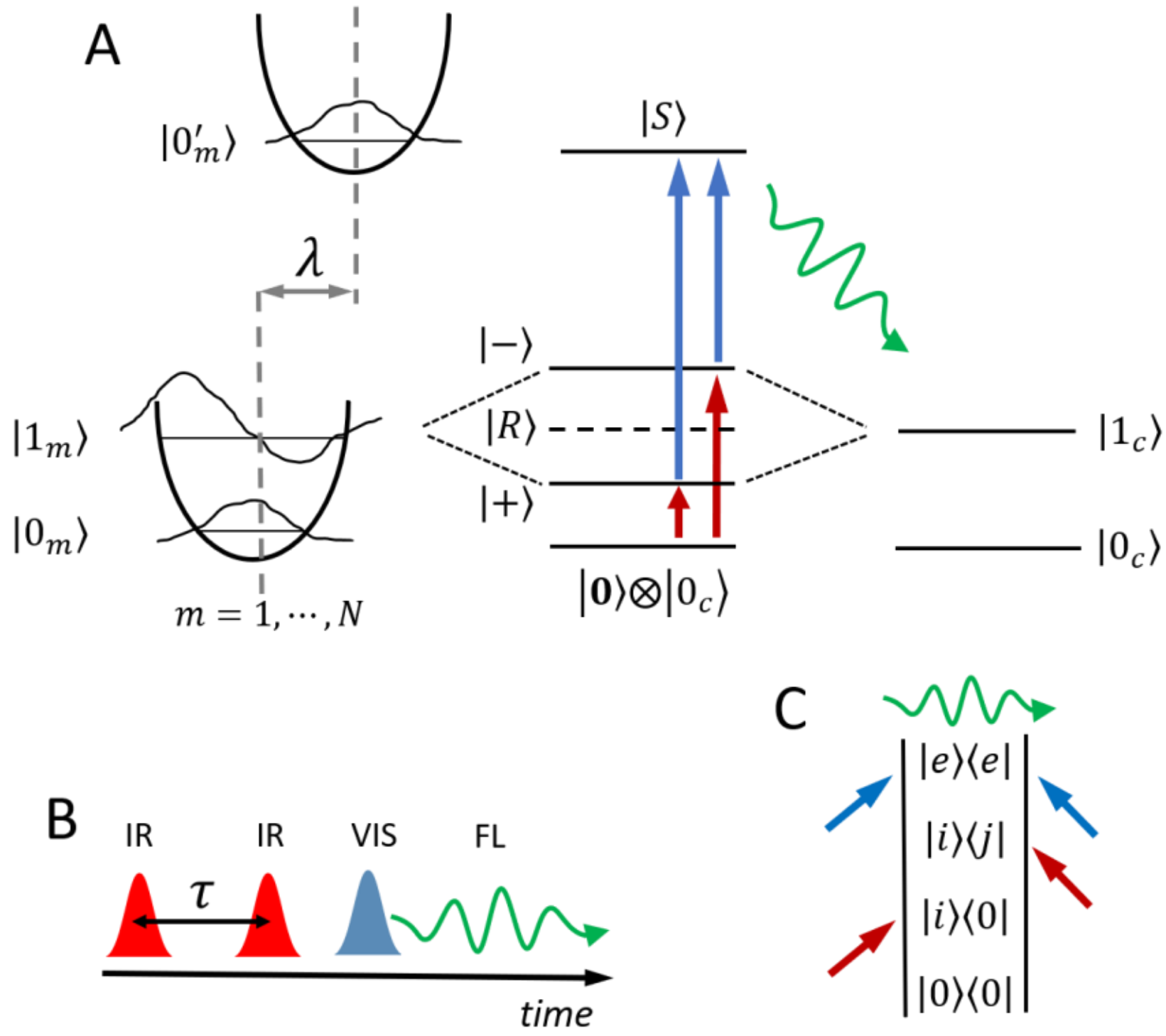

**Fig. 1. FE-IR spectroscopy of VSC system.** (A) Excitation states diagram. Left: ground and excited molecular electronic states displaced by the vibronic constant $\lambda$ along the normal mode coordinate. The molecules in the ensemble enumerated by index $m$ contribute to collective bright state $|B\rangle$. Right: photonic states of infrared cavity. Middle: vibro-polariton states of the hybrid system. Red arrows – infrared laser pulses, blue arrows – visible laser pulses, green arrow – fluorescence emission. (B) Sequence of excitation laser pulses, $\tau$ – coherence time interval. (C) Double-sided Feynman diagram describing the Liouville space fourth-order response functions. Indices $i, j$ - vibrational (infrared cavity) excitations, $e$ - excited electronic state. Horizontal arrow on top of the diagram - fluorescence emission.

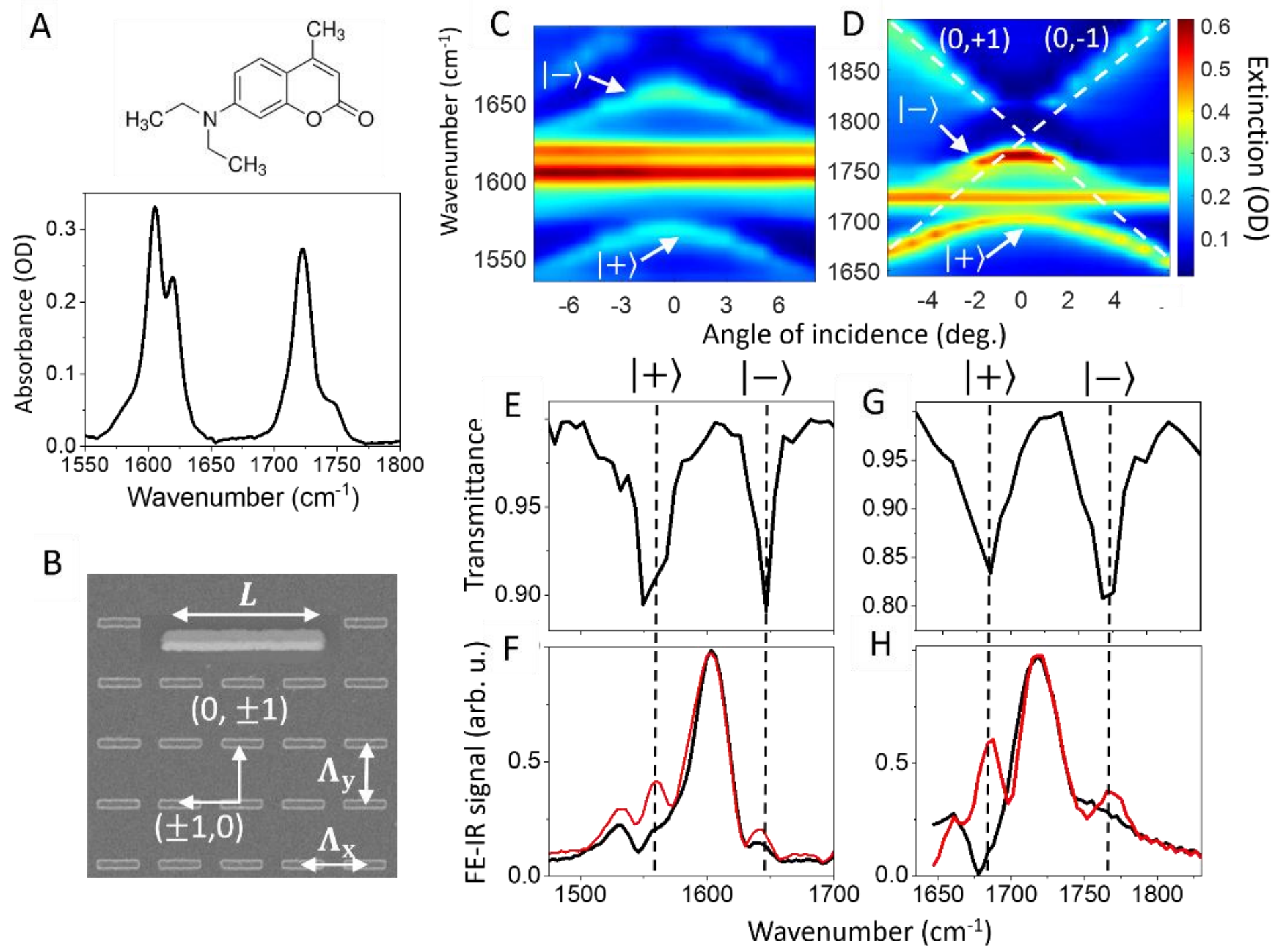

**Fig. 2. VSC system.** (A) Linear FT-IR spectrum of C-1 solution in toluene. Molecular structure is shown on top. (B) Scanning electron micro-image of the infrared antenna array. Antenna dimensions are: length ($L$) 1.1 μm, width 200 nm, and height 80 nm. Lattice diffraction orders associated with SLRs are indicated. Vibro-polaritons dispersion maps obtained with FT-IR. (C) C=C vibration strongly coupled to SLR. Spectra collected at different angles of incidence with TE-polarized light. Array lattice constants are $\Lambda_x$=1.6 μm and $\Lambda_y$=4.2 μm. Polariton branches are marked with white arrows. (D) Same as panel C, but for C=O vibration. Array lattice constants are $\Lambda_x$=1.6 μm and $\Lambda_y$=3.8 μm. Non-linear FE-IR spectroscopy. (E) Transmission spectrum for near-normal incidence of IR laser pulses. The bare-molecule spectrum was used as background. Vibro-polaritons formed with C=C modes indicated with vertical dashed lines. (F) FE-IR spectrum of the bare-molecule (black) and of the VSC molecules (red). (G) Same as in panel E, but for C=O modes. (H) Same as in panel F, but for C=O modes.

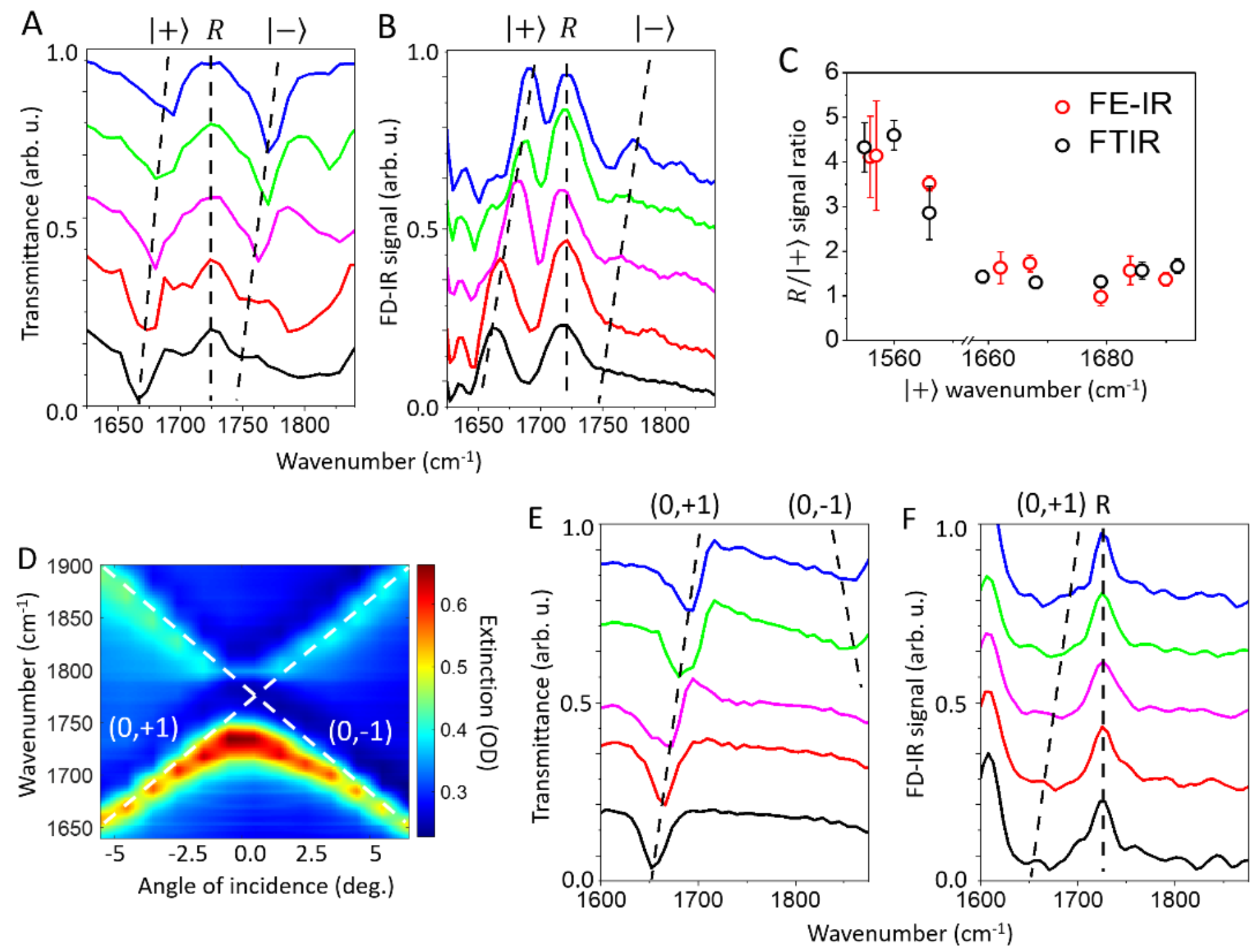

**Fig. 3. FE-IR spectroscopy of C=O vibration on antenna arrays.** FE-IR measurements with different incident angles of IR pulses. Blue - $\theta \sim 1°$, green - $\theta \sim 2°$, magenta - $\theta \sim 4°$, red - $\theta \sim 5°$, black - $\theta \sim 6°$. Polariton and reservoir peaks are traced by dashed lines. VSC regime, C-1 concentration 2.5 M. (A) Transmission spectra measured with bare-molecule spectrum as background. (B) FE-IR spectra. (C) Reservoir to $|+\rangle$ signal ratio obtained with FT-IR (black circles) and FE-IR (red circles). The data are plotted for different $|+\rangle$ transition frequencies. Uncoupled C=O vibration, C-1 concentration 40 mM. (D) SLR dispersion map measured with FT-IR. (E) Transmission spectra, measured with bare-molecule spectrum as background. (F) FE-IR spectra. SLR and bare-molecule transitions frequencies are traced with dashed lines.

**Supplementary Information**

**Supplementary Text**

**The Feshbach projection operator formalism for $\widehat{\widetilde{H}}_{\text{tot}}$**

To estimate the scaling of effective coupling between the cavity and the electronic excitations in the presence of the laser field, we consider the driven TCH model in the space spanned by $|c\rangle$, $|S\rangle$, and $|B\rangle$:

$$\widehat{\widetilde{H}}_{\text{tot}} = \hbar\omega_c|c\rangle\langle c| + \hbar\widetilde{\omega}_e|S\rangle\langle S| + \hbar\omega_v|B\rangle\langle B| + \left[g_{\text{tot}}|c\rangle\langle B| + \mu_{1,0'}E_L(t)e^{i\omega_L t}|S\rangle\langle B| + h.c.\right].$$

Any instantaneous eigenstate of this Hamiltonian can be projected onto the subspace spanned by $|c\rangle$ and $|S\rangle$, using the Feshbach's projectors procedure. The effective Hamiltonian is given by

$$\widehat{H}_{\text{eff}}(E) = QHQ + QHP\frac{1}{E - PHP + i\varepsilon}PHQ,$$

where $E$ is the transient eigenvalue of $\widehat{\widetilde{H}}_{\text{tot}}$, and the projectors $P$ and $Q$ make the complete basis set: $Q = |c\rangle\langle c| + |S\rangle\langle S|$, $P = |B\rangle\langle B|$ and $P + Q = \boldsymbol{I}$. Setting $E_L(t) \approx$const we obtain,

$$\widehat{H}_{\text{eff}}(E) = \left(\hbar\omega_c + \frac{g_{\text{tot}}^2}{E - \hbar\omega_v + i\varepsilon}\right)|c\rangle\langle c| + \left(\hbar\widetilde{\omega}_e + \frac{1}{E - \hbar\omega_v + i\varepsilon}\right)|S\rangle\langle S| + \left[\frac{g_{\text{tot}}\mu_{1,0'}E_L e^{i\omega_L t}}{E - \hbar\omega_v + i\varepsilon}|c\rangle\langle S| + c.c.\right].$$

The coupling (off-diagonal) term is shown to scale as the product of the VSC strength ($g_{\text{tot}}$) and the laser induced transition strength ($\mu_{1,0'}E_L e^{i\omega_L t}$), indicating the necessity of the laser field in transforming the vibrational strong coupling in the infrared into a collective electronic excitation.

**Supplementary figures**

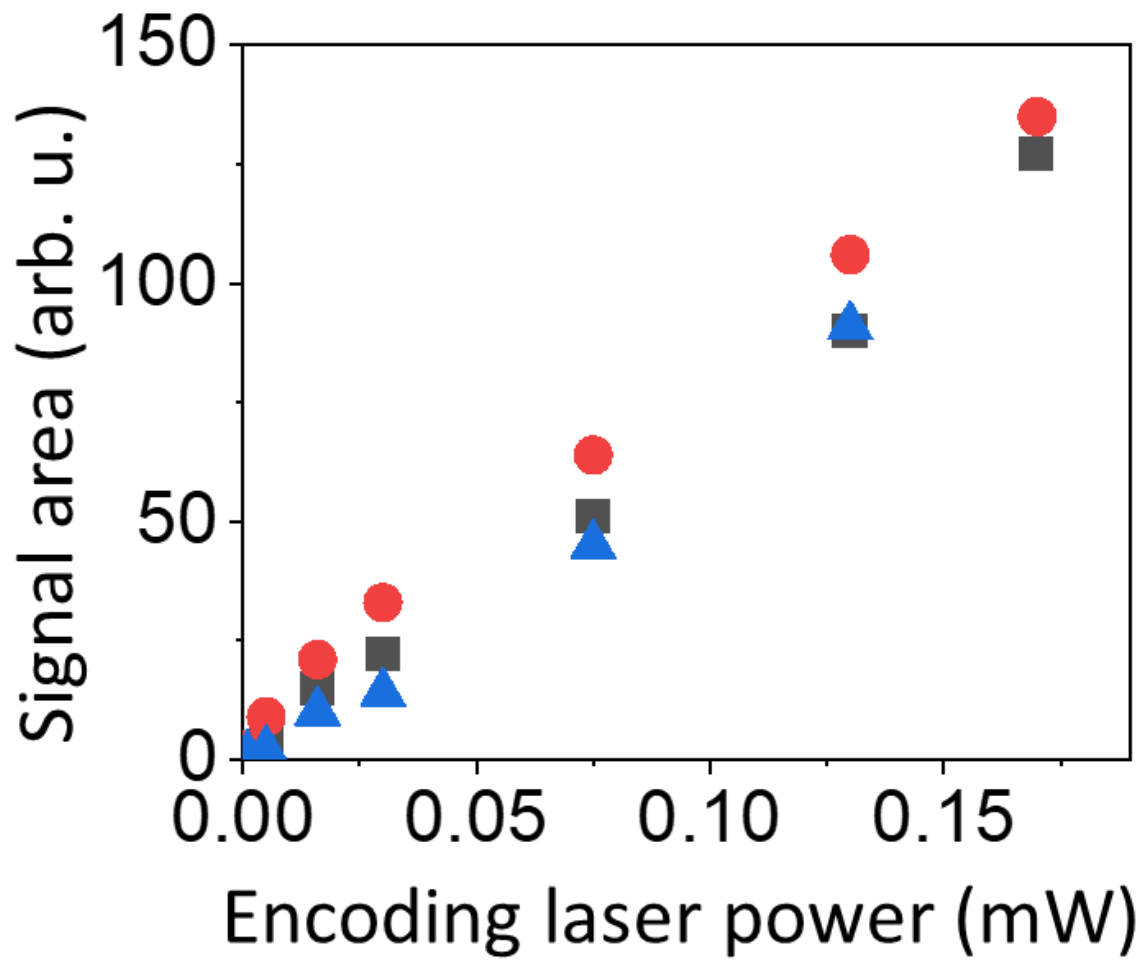

**Fig. S1. Encoding laser power dependence of FE-IR signals in VSC system.** Red circles – reservoir, blue triangles - $|+\rangle$ state, black squares - $|-\rangle$ state.